\documentclass[preprint]{revtex4}
\usepackage{epsf}

\newlength{\messboxwidth}
\newcommand{\messbox}[2]{\vskip 5mm\noindent
\fbox{\parbox{\messboxwidth}{{\bf Lesson #1:} #2}}\vskip 5mm}

\def\kB{k_{\rm B}}

\begin{document}
\bibliographystyle{apsrev}

\title{Systematic coarse graining: ``Four lessons and a caveat''
from nonequilibrium statistical mechanics}

\author{Hans Christian \"Ottinger}
\email[]{hco@mat.ethz.ch}
\homepage[]{http://www.polyphys.mat.ethz.ch/} \affiliation{ETH
Z\"urich, Department of Materials, Polymer Physics, HCI H 543,
CH-8093 Z\"urich, Switzerland}

\begin{abstract}
With the guidance offered by nonequilibrium statistical thermodynamics, simulation techniques are elevated from brute-force computer experiments to systematic tools for extracting complete, redundancy-free and consistent coarse grained information for dynamic systems. We sketch the role and potential of Monte Carlo, molecular dynamics and Brownian dynamics simulations in the thermodynamic approach to coarse graining. A melt of entangled linear polyethylene molecules serves us as an illustrative example.
\end{abstract}

\maketitle

\section*{Goals}
The goal of coarse graining is to minimize the
efforts required to solve a given problem. It moreover puts the
focus on the essence of the problem and hence maximizes the depth
of understanding. As coarse graining is the key to both efficiency
and understanding, reliable general principles and procedures need
to be established. We here summarize and illustrate a systematic
approach based on nonequilibrium statistical mechanics. This theme
article provides some useful guidelines for physically meaningful
coarse graining. The author hopes to be thought-provoking, which
may be more realistic for young scientists but equally important
for experienced colleagues who have already carried out
nonequilibrium simulations.

\section*{Beyond Equilibrium Thermodynamics}
At equilibrium, statistical thermodynamics
provides universally accepted recipes for coarse graining: (i)
Thermodynamic potentials contain all the relevant information
about thermodynamic systems in a compact format, and (ii)
thermodynamic potentials can be calculated via partition
functions. The ultimate task of computer simulations is to
evaluate partition functions to obtain complete and consistent
thermodynamic information. Simulations should not be regarded as
computer experiments in which certain equations of state are
``measured;'' otherwise, one is at risk to violate the Maxwell
relations between different equations of state instead of
benefiting from the simplifications offered by these thermodynamic
relationships.

Beyond equilibrium, we need to identify the generalizations of
thermodynamic potentials as the primary sources of complete
thermodynamic information. In response to this need, GENERIC
(``general equation for the nonequilibrium reversible-irreversible
coupling'') offers a complete, coherent, and consistent
theoretical framework of beyond-equilibrium thermodynamics. It is
important to consider the energy $E$ and the entropy $S$
separately because they are the distinct generators of reversible
and irreversible dynamics, respectively. In addition to the two
generators $E$ and $S$, one needs a Poisson matrix $L$ that turns
energy gradients into reversible dynamics and a friction matrix
$M$ that turns entropy gradients into irreversible dynamics. The
fundamental equation for the time evolution of the nonequilibrium
variables $x$ is
\begin{equation} \label{LMformulation}
  \frac{dx}{dt} = L(x) \cdot \frac{\delta E(x)}{\delta x} +
  M(x) \cdot \frac{\delta S(x)}{\delta x} .
\end{equation}
This equation is known as GENERIC \cite{hco99,hco100,hcobet}. It
comes with a number of structural properties of the generators
$E$, $S$ and the matrices $L$, $M$. For example, $L$ is always
antisymmetric and $M$ is usually symmetric (more precisely, $M$
actually expresses the famous Onsager-Casimir symmetry of linear
irreversible thermodynamics \cite{hcobet,deGrootMazur}). A strict
separation of the reversible and irreversible contributions is
provided by the mutual degeneracy requirements
\begin{equation} \label{consistency}
  M(x) \cdot \frac{\delta E(x)}{\delta x} = 0 , \qquad
  L(x) \cdot \frac{\delta S(x)}{\delta x} = 0 ,
\end{equation}
which are strong formulations of the conservation of energy even
in the presence of dissipation and the conservation of entropy for
any reversible dynamics. Entropy can only be produced by
irreversible dynamics. A strong nonequilibrium generalization of
the second law of thermodynamics is obtained by postulating that
$M$ be positive-semidefinite.

Because $x$ typically contains position-dependent fields, such as
the local mass, momentum and energy densities of hydrodynamics,
the state variables are usually labeled by continuous (position)
labels in addition to discrete ones. It is important to realize
that the choice of the variables $x$, that is, the definition of
an appropriate thermodynamic system for a given problem, is a step
of crucial importance. Beyond equilibrium, and contrary to our
equilibrium experience, this choice is far from obvious. Actually,
this is the point at which the most physical intuition is
required. A poor choice of $x$ cannot be repaired by even the most
ingenious formulation of thermodynamic building blocks. For
example, for the famous reptation model of a melt of entangled
linear polymer molecules considered below, the idea of smooth
primitive paths and the corresponding configurational distribution
function are the key to success.

\section*{Beyond Equilibrium Statistical Mechanics}
The essence of nonequilibrium statistical
mechanics and coarse graining is contained in Figure~1 for the
entropy $S$ and Figure~2 for the friction matrix $M$. The entropy
on the coarser level is obtained by counting the increasing number
of microstates lumped into coarser states for progressive coarse
graining, and the friction matrix is associated with the
increasing number of processes treated as fluctuations upon
restricting the coarser description to slower and slower variables
(according to the fluctuation-dissipation theorem, which holds
even far beyond equilibrium, fluctuations are intimately related
to dissipation or friction).

To classify and count microstates in the spirit of Figure~1, it is
crucial to have a mapping $\Pi(z)$ that assigns a coarse grained
state to any microstate $z$. In classical mechanics, the
microstates $z$ are given by the positions and momenta of all
particles. In a generalized microcanonical ensemble, all
microstates $z$ associated with the same coarse configuration
$\Pi(z)$ occur with the same probability so that the mapping
$\Pi(z)$ fully characterizes the probability density $\rho_x(z)$
to find a microstate $z$ for given coarser variables $x$. In any
ensemble, these variables $x$ are the averages of $\Pi(z)$
evaluated with the probability density $\rho_x(z)$, i.e.\ $x =
\langle\Pi(z)\rangle_x$.

Like at equilibrium, it is convenient to pass from microcanonical
to canonical ensembles to simplify practical calculations. One
then controls the averages of $\Pi(z)$ to be $x$ (or some of them)
by a corresponding set of Lagrange multipliers $\lambda$. The
passage from one ensemble to another one works exactly as in
equilibrium statistical mechanics, and it is closely related to
Legendre transformations. The only difference is that the
nonequilibrium list of variables is usually much longer and less
universal than the list of equilibrium variables. For a larger
list of variables, the number of microstates per coarse grained
state goes down, so that the statistical equivalence of different
ensembles needs to be investigated more carefully than at
equilibrium.

The concrete statistical recipe behind Figure~1 is that the
entropy of a coarse state is given by the logarithm of the number
of associated microstates. Figure~1 visualizes the pattern
recognition process behind lumping microstates into coarser states
in a simplified cartoon. In general, a suitable representation
needs to be identified before any clustering of microstates can be
recognized, which is again the art of finding suitable variables.
Figure~2 illustrates the essence of the friction matrix and
actually represents the formula
\begin{equation}\label{GreenKubo}
  M(x) = \frac{1}{\kB} \int_0^\tau \left \langle
  \dot{\Pi}^{\rm f}(z(t)) \dot{\Pi}^{\rm f}(z(0)) \right \rangle_x dt ,
\end{equation}
where $\kB$ is Boltzmann's constant, $\tau$ is an intermediate
time scale separating the slow degrees of freedom from the fast
degrees of freedom, $\dot{\Pi}^{\rm f}(z)$ is the rapidly
fluctuating part of the rate of change of the atomistic
expressions $\Pi(z)$ for the slow variables $x$, and the average
is over an ensemble of atomistic trajectories $z(t)$ consistent
with the coarse grained state $x$ at $t=0$ and evolved according
to the fast contribution to the atomistic dynamics to the time
$t$. Equation (\ref{GreenKubo}) is known as a Green-Kubo formula
\cite{KuboetalII,Grabert}.

The GENERIC building blocks $E$ and $L$ associated with reversible
dynamics are obtained by straightforward averaging of the
microscopic energy and of the Poisson bracket of classical
mechanics for the components of $\Pi(z)$. We thus have a complete
set of recipes to calculate the four GENERIC building blocks of
nonequilibrium thermodynamics from statistical mechanics. All the
building blocks and their properties have been derived from
Hamilton's equations of motion by separating the slow and fast
degrees of freedom with the projection-operator method
\cite{hcobet,Grabert,hco101}. As already suggested by Figures~1
and 2, the statistical recipes can be generalized to progressive
coarse graining rules so that it is not necessary to start always
from the atomistic level. Time occurs only in the expression
(\ref{GreenKubo}) for the friction matrix $M$; all other building
blocks are obtained by time-independent averaging or counting
according to the statistical ensemble. All dynamic material
information must hence be contained in $M$.

We conclude this section on nonequilibrium statistical mechanics
with a warning that is obvious within the current approach
(remember Figures~1 and 2), but is often ignored in the literature
on coarse graining:

\vskip 5mm\noindent \fbox{\parbox{\messboxwidth}{{\bf Caveat:} The
hallmark of true and complete coarse graining is an increase of
both entropy and dissipation (or friction); meaningful coarse
graining techniques focus on calculating both increases or, at
least, are capable of accounting for the corresponding
effects.}}\vskip 5mm

The passage from reversible to irreversible equations is one of
the big intellectual achievements associated with Boltzmann's
kinetic equation for rarefied gases. Such a transition from
reversible to irreversible (or, the emergence of a nonzero
friction matrix) must occur in any coarse graining starting from
the reversible atomistic level. Even when starting from a level
with an irreversible contribution to the dynamics, additional
irreversibility must be born in a coarse graining process
(otherwise, we have a solution technique without any elimination
of fast processes in favor of fluctuations). The issue of coarse
graining versus reduction (i.e., solving) is a hot topic in
mathematical physics \cite{hcoredcg}.

In many cases, one tries to coarse grain systems by introducing
temperature-dependent effective potentials (more precisely: free
energies) for ``superatoms'' which typically consist of some $10$
atoms. To calculate such effective potentials efficiently,
``iterative Boltzmann inversion'' has been developed. With this
approach, one can then handle the additional entropy
appropriately, but not the additional dissipation, the emergence
of (further) irreversibility. In many cases, the effective
potentials are simply used in the reversible equations of motion
of classical mechanics to perform molecular dynamics simulations
(see \cite{MilanoMPla05,SunFaller06} and references therein). A
simplistic attempt to allow at least for some kind of dissipative
effect would be to renormalize time
\cite{MilanoMPla05,Harmandetal06} or to introduce frictional
forces into the equations of motion \cite{PaddingBriels02}. The
example of hydrodynamic interactions in polymer solutions,
however, warns us that the non-scalar nature of the friction
matrix $M$ is crucial to find many of the relevant physical
effects such as second-normal stress differences in shear flow
\cite{hco23,hco33}. The entire wealth of dynamic phenomena is
actually contained in $M$, so that a simple rescaling of time or a
scalar friction can only be of limited value.

\section*{The Lessons (for Beginners and Professionals)}
It is worthwhile to emphasize once more that
identifying the proper relevant variables is the key to success in
any systematic coarse graining method. The initial step is hence
clear:

\messbox{1}{Identify a suitable set of relevant variables for a
coarser target level and express them in terms of the variables of
a finer source level.}

The finer level is considered to be well-established. In
many cases, it is the atomistic level with particle positions and
momenta as variables. In some cases, the art of finding
appropriate force fields is supported by quantum calculations.

The implications of ``Lesson~1'' are dramatic: We cannot start a
simulation before we have found suitable coarse grained variables,
that is, before we have reached a basic understanding of our
problem of interest. This is the price to pay for the possibility
to do systematic and iterated coarse graining to bridge a wide
range of scales. In brute-force computer experiments, the system
is driven by boundary conditions, nonequilibrium initial
conditions, and/or external forces chosen to mimic the
experimental situation of interest. One then ``measures'' some
information of direct interest. In statistical mechanics, the
system is considered in terms of well-defined nonequilibrium
ensembles and they involve the coarse grained variables
explicitly.

In the presence of self-similarity or universality (or in the
absence of better ideas), one takes a target level of the same
type as the source level. This is, for example, the idea behind
``superatoms'' (without any justification by self-similarity or
universality). In general, ingenious ideas about suitable target
levels promise a more rewarding starting point for significant
coarse graining. A variety of inspiring ideas has been collected
in a recent monograph by Kr{\"o}ger \cite{Kroeger}.

The choice of variables determines the observables to be averaged
to obtain the thermodynamic building blocks in which the complete
coarse grained information is collected. Moreover, the form of the
statistical ensemble depends strongly on the chosen variables.

\messbox{2}{Choose a convenient nonequilibrium ensemble, verify
its equivalence with a generalized microcanonical ensemble,
interpret all the Lagrange multipliers, and find their values for
nonequilibrium situations of interest.}

In general, the identification of the proper values of
the Lagrange multipliers requires an iterative procedure. For
example, the Lagrange multipliers required to describe the
structure of a complex fluid undergoing homogeneous steady flow
can only be related to the velocity gradients if the friction
matrix $M$ is known \cite{hco146} (from an iterative procedure
leading to self-consistency). The highly advantageous restriction
to homogeneous situations is often possible due to the local
character of the thermodynamic building blocks. Biased Monte Carlo
methods, parallel tempering \cite{MarinParisi92,Tesietal96},
multicanonical methods \cite{FallerYandePab02} and
density-of-states Monte Carlo
\cite{WangLandau01a,WangLandau01b,FallerdePablo03} offer
interesting possibilities for locating self-consistent values of
Lagrange multipliers efficiently.

\messbox{3}{Calculate the static building blocks $E$, $S$, and $L$
of nonequilibrium thermodynamics by the highly developed Monte
Carlo methods of equilibrium statistical mechanics
\cite{LandauBinder,BinderHeerm,Mouritsen}. Overcome the challenges
implied by a wide range of scales in a problem by a balanced set
of Monte Carlo moves causing configurational changes occurring on
various length and time scales.}

In more provocative words, the most unphysical Monte Carlo moves,
such as breaking and recombining polymer chains, are most
promising for overall computational efficiency. Whenever Monte
Carlo time can be assumed to be proportional to real time, this is
a guarantee for computational inefficiency.

In many cases, the Poisson matrix $L$ can be obtained analytically
from the transformation behavior of the variables under space
transformations without performing any simulations. Then, only
energy and entropy need to be found from Monte Carlo simulations.

\messbox{4}{Evaluate the friction matrix $M$ from the Green-Kubo
formula (\ref{GreenKubo}). As the only building block containing
dynamic material information, its calculation requires dynamic
simulation techniques. However, dynamic simulations are required
only for a small fraction of the longest time scales of the
system. The only goal of long-time stabilization is to conserve
the chosen statistical ensemble.}

If one starts from the atomistic level, molecular
dynamics simulations \cite{AllenTildes} are the natural tool.
Starting from a coarse grained level, for which fast processes are
treated as white noise, the equations of motion become stochastic
differential equations and Brownian dynamics simulations are
required \cite{hcobook}.

It is often stated that the reliable determination of dynamic
material properties such as the diffusion coefficient requires
trajectories that are large compared to the longest relaxation
time of the system. Statistical mechanics tells us that all the
relevant information is actually obtained within a fraction of a
relaxation time. The Green-Kubo formula (\ref{GreenKubo}) for the
friction matrix, which is the only thermodynamic building block
containing dynamic material information, uses the intermediate
time scale $\tau$ as the upper limit of integration. This time
scale is long from the perspective of the fast processes treated
as fluctuations (correlations of fluctuations decay to zero within
$\tau$) and short from the perspective of the relevant slow
processes (they are arrested). Typically, the longest relaxation
time and the intermediate time scale $\tau$ are not known in
advance. We hence try to estimate the friction matrix from
increasingly longer trajectories until stable results are
obtained, which typically determine the characteristic relaxation
times of a system. We can then check whether the required length
of the trajectory is much shorter than the longest relaxation
times so that there actually exists a separating time scale $\tau$
inbetween.

Taking ``Lesson~4'' to heart typically is rewarded with two orders
of magnitude in computational efficiency. This miracle happens
because we do not need to reveal the relevant relaxation
mechanisms in thermodynamically guided simulations; they are
implied by the choice of variables, emphasizing once more how
important this initial choice (``Lesson~1'') is. To check the
adequacy of the choice of variables, we should employ the coarse
grained level to make predictions for a variety of specific
situations, which we can compare to simulations on the more
detailed level or to experiments.

In most dynamic simulations, one tries to produce very long
trajectories. The practical advantage is that one then does not
need any separate Monte Carlo simulations. From the viewpoint of
efficiency, however, one loses the benefit of ``Lesson~3'' that
Monte Carlo can beat the time scale problem. If one performs these
long dynamic simulations to avoid Monte Carlo, one has to
stabilize the long-time dynamics, for example by thermostats, to
conserve the chosen statistical ensemble (see Section 8.4.3 of
\cite{hcobet} and references therein). For that purpose, the ideas
behind the Nos\'e-Hoover \cite{Nose84,Hoover95} and
Nos\'e-Poincar\'e \cite{BondLeimkLaird99,Leimkuhler02} thermostats
as well as Andersen's barostat \cite{Andersen80} are particularly
useful. Any modification of the atomistic equations of motion is
allowed as long as it has no significant influence on the dynamics
on time scales shorter than $\tau$, that is, on the dynamic
material information. Actually, also in the formal
projection-operator results, the slow degrees of freedom are
projected out of the time evolution in the Green-Kubo formula.

\section*{Example: Entangled Polyethylene}
In the spirit of ``Lesson~1'', we need a
suitable coarse grained structural variable before we can start a
thermodynamically guided simulation technique. Our choice for
entangled linear polymer chains is motivated by the famous
reptation model of Doi and Edwards \cite{DoiEdwards}. From the
random coil conformations of the atomistic polymer chains we
construct smooth primitive paths or worms (see Figure~3) which
allows us to define a tangent unit vector $\bbox{u}$ at any
position $\sigma$ along the chain ($0 \leq \sigma \leq 1$)
\cite{hco142}. The probability density $f(\bbox{u},\sigma)$ serves
as our structural variable.

The corresponding Lagrange multiplier $\lambda(\bbox{u},\sigma)$
in a generalized canonical ensemble (``Lesson~2'') can be
identified as an ``effective potential'' characterizing the
deviation from the equilibrium probability density via $f = f_{\rm
eq} e^{-\lambda}$, where $f_{\rm eq}$ is just a constant for the
reptation model. The simplicity of this relationship is a
consequence of the simple form of Boltzmann's entropy for
probability densities. With the chosen structural variable, we
fully rely on the information about the orientation $\bbox{u}$ of
a single segment $\sigma$ without any correlation between
different segments or different chains. Then, the stiffness of the
worms enters as an external parameter equivalent to the number of
entanglements per chain. To avoid that, one would need to elevate
the number of entanglements to the level of a dynamic variable.
After borrowing the ingenious structural variable
$f(\bbox{u},\sigma)$ from Doi and Edwards and choosing a
convenient ensemble, the way to thermodynamically guided
simulations is straightforward. No further intuition is required.
More details on the example sketched in the following can be found
in Section 8.4.6 of \cite{hcobet}.

Figure~3 suggests naturally that, in addition to the contour of
the worms, one might consider the worm cross section as a relevant
variable. Such a modification of the Doi-Edwards model has
previously been suggested to explain the experimental results for
the second normal-stress difference in shear flow
\cite{IanniMarrucci98,MarrucciIanni99,hco119}.

Even before we start a simulation, a number of further conclusions
can be drawn from closer inspection of the statistical expressions
for the thermodynamic building blocks. There is no energy
associated with the variable $f(\bbox{u},\sigma)$, and the entropy
associated with any probability density is expected to be of the
Boltzmann type. Therefore, for the single-segment reptation model,
the stresses in a polymer melt are entirely entropic in nature.
Note that such a statement depends on the level of description. On
the atomistic level, the very same stresses are obtained from the
interactions between all atoms of all chains.

To construct the Poisson matrix, we only need to know that the
orientation vector $\bbox{u}$ is rotated by the flow field. The
only quantity to be determined from simulations is the friction
matrix. As all static building blocks are known, there is no need
for Monte Carlo simulations (``Lesson~3''). We here determine the
friction matrix at equilibrium, that is, for $\lambda = 0$, by
means of a molecular dynamics simulation with a well-established
force field for polyethylene
\cite{HarmMavTheo98,HarmMavTheo00,hco145} (``Lesson~4''). The
structural variable then does not occur in the definition of the
ensemble, but it still determines the correlation functions to be
considered in the Green-Kubo formula (\ref{GreenKubo}). Analytical
rearrangements of the Green-Kubo formula show that we need to find
second moments of displacements of $\sigma$ and $\bbox{u}$ up to
intermediate time scales, short compared to the longest relaxation
time which is also known as the disengagement or reptation time.

Orientational diffusion of $\bbox{u}$ has actually been proposed
as a ``constraint-release mechanism'' \cite{hco74} that is absent
from the original Doi-Edwards model. We here focus on the
diffusion of $\sigma$, which is the diffusion of worms along their
contours and thus the classical reptation mechanism. The fact that
we need to focus on the reptative motion of the worms is a
straightforward result of the thermodynamically guided approach
carried out for the structural variable $f(\bbox{u},\sigma)$.
There is no need to look at any other dynamic quantity, such as
the stress tensor or center-of-mass diffusion, unless one wants to
investigate the adequacy of the coarse grained level.

Figure~4, which is based on data kindly provided by Vlasis
Mavrantzas, shows the average square displacement of $\sigma$ as a
function of time for polyethylene chains with $1000$ carbon atoms
per chain. The slope in this figure determines the frictional
properties and hence the disengagement time, which is obtained as
$(6.8 \pm 0.2)\,\mu$s. Note that this result is obtained from a
trajectory of less than $0.5\,\mu$s. This is a striking example of
``Lesson~4'': The frictional properties on the coarse grained
level, which determine the disengagement time, can be obtained
reliably on a much shorter, intermediate time scale. Total
disengagement requires $(\Delta_t \sigma)^2$ to be of order unity;
it is hence clear from Figure~4 that the disengagement time is
indeed determined from the early traces of reptation becoming
visible through the apt choice of smooth primitive paths as
variables. To obtain the same material information from the
center-of-mass diffusion, one needs trajectories comprising
several relaxation times. This gain of efficiency hopefully is a
convincing example for the benefits of learning the lessons of
nonequilibrium statistical mechanics. Equally important is the
gain of understanding provided by the reptation model of Doi and
Edwards, which enables the computational efficiency of systematic
coarse graining.

In conclusion, the best strategy for thermodynamically guided,
systematic coarse graining requires a combination of Monte Carlo
simulations for the efficient calculation of energy and entropy
and dynamic simulations for the friction matrix. Monte Carlo
simulations can also provide the initial conditions for dynamic
simulations. Stand-alone Monte Carlo simulations are usually of
limited value for nonequilibrium systems because the
identification of the Lagrange multipliers describing deviations
from equilibrium requires dynamic material information
\cite{hco146}. In spite of the rapidly developing computer
technology and the increasing temptation to solve all engineering
problems by brute-force computer experiments, the deeper
understanding required and enriched by systematic coarse graining
will hopefully be appreciated in perpetuity.

\section*{Acknowledgments}
I thank Juan Jos\'e de Pablo for inviting and
inspiring me to write this theme article. My long-standing,
enjoyable and fruitful collaboration with Vlasis Mavrantzas and
Martin Kr\"oger  contributed enormously to the development of
systematic coarse graining procedures. I greatly appreciate Dieter
Schl\"uter's constructive comments which helped to intensify my
messages.

\newpage
\section*{Figures}
\vspace{1cm}
\centerline{\epsfxsize=10cm \epsffile{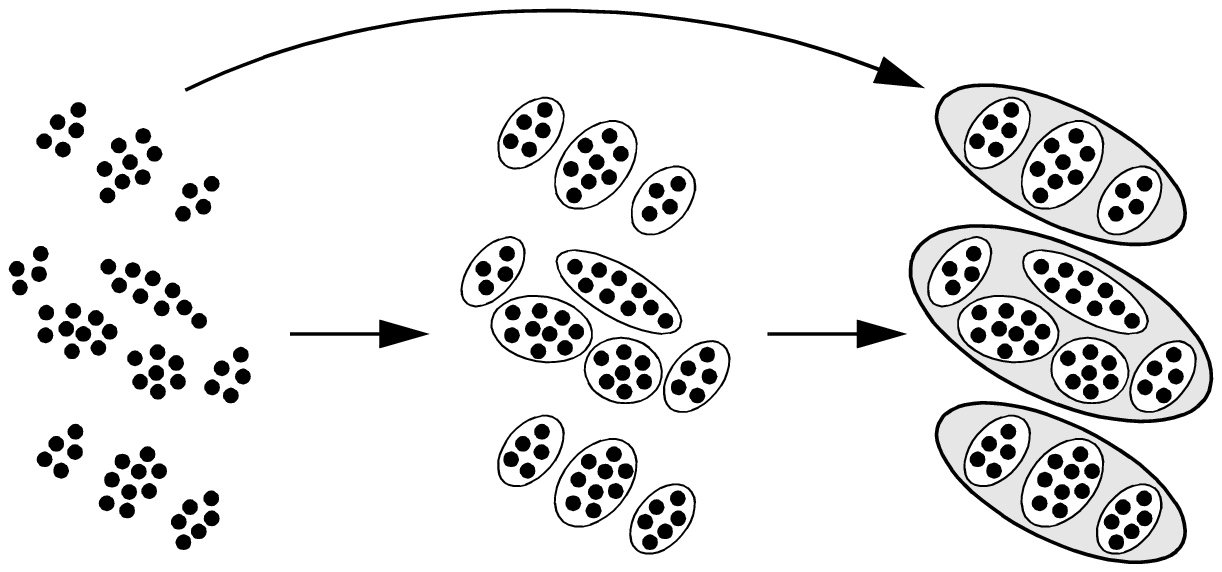}}
\vspace{1.5cm}

\noindent\emph{Figure 1. Grouping of microstates (black circles) into
progressively coarser states (ellipses); following
well-established equilibrium ideas, the nonequilibrium entropy of
a coarse state is given by the logarithm of the number of
microstates in the corresponding ellipse.}

\vspace{4cm}
\centerline{\epsfxsize=9cm \epsffile{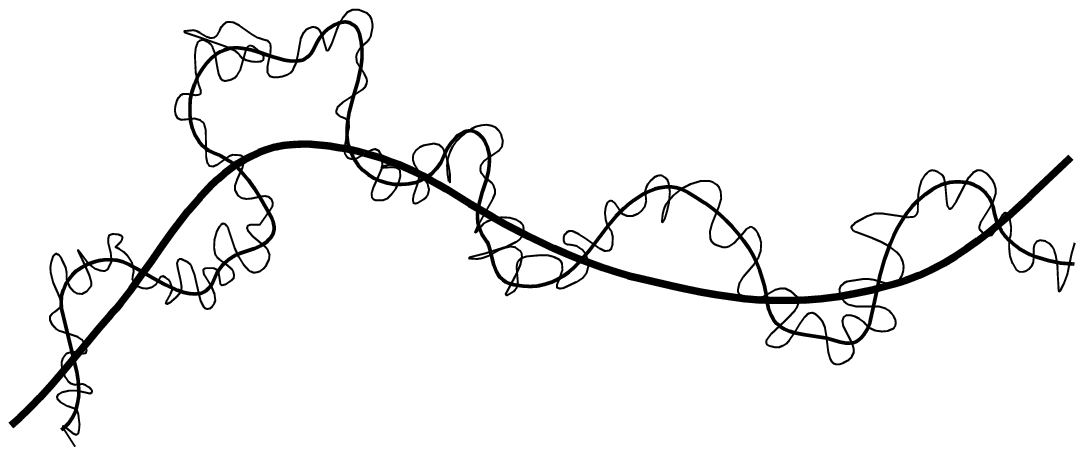}}
\vspace{1.5cm}

\noindent\emph{Figure 2. Trajectories on three levels of
description; an increasing number of fast processes are eliminated
in favor of random fluctuations (not shown in the figure) for the
coarser descriptions (for increasing line thickness).}

\newpage
\vspace*{.1cm}
\centerline{\epsfxsize=9cm \epsffile{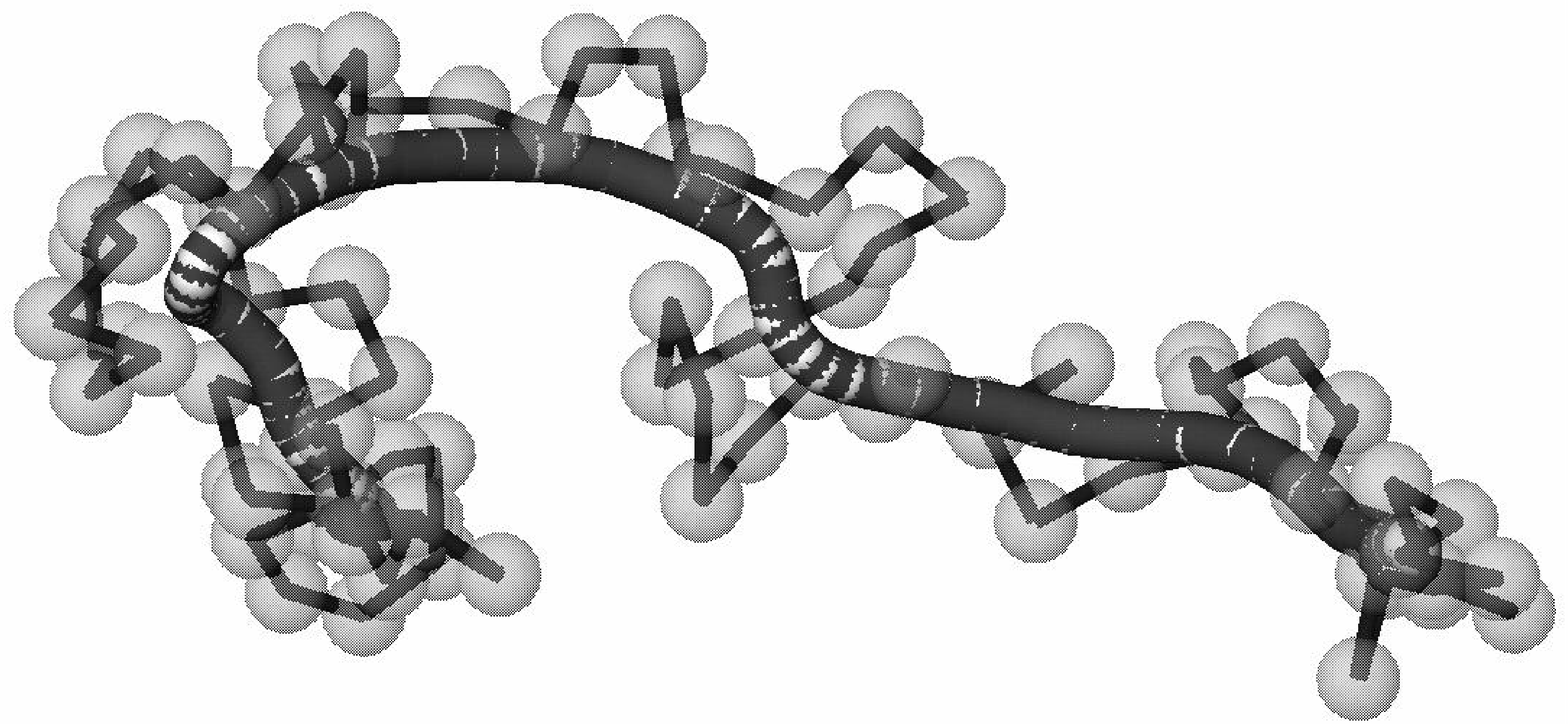}}
\vspace{1.5cm}

\noindent\emph{Figure 3. Competing-spring construction of a smooth
primitive path from an atomistic chain; the smoothed worm
possesses a local tangent vector. [Figure courtesy of M.~Kr\"oger,
ETH Z\"urich.]}

\vspace*{3cm}
\centerline{\epsfxsize=9cm \epsffile{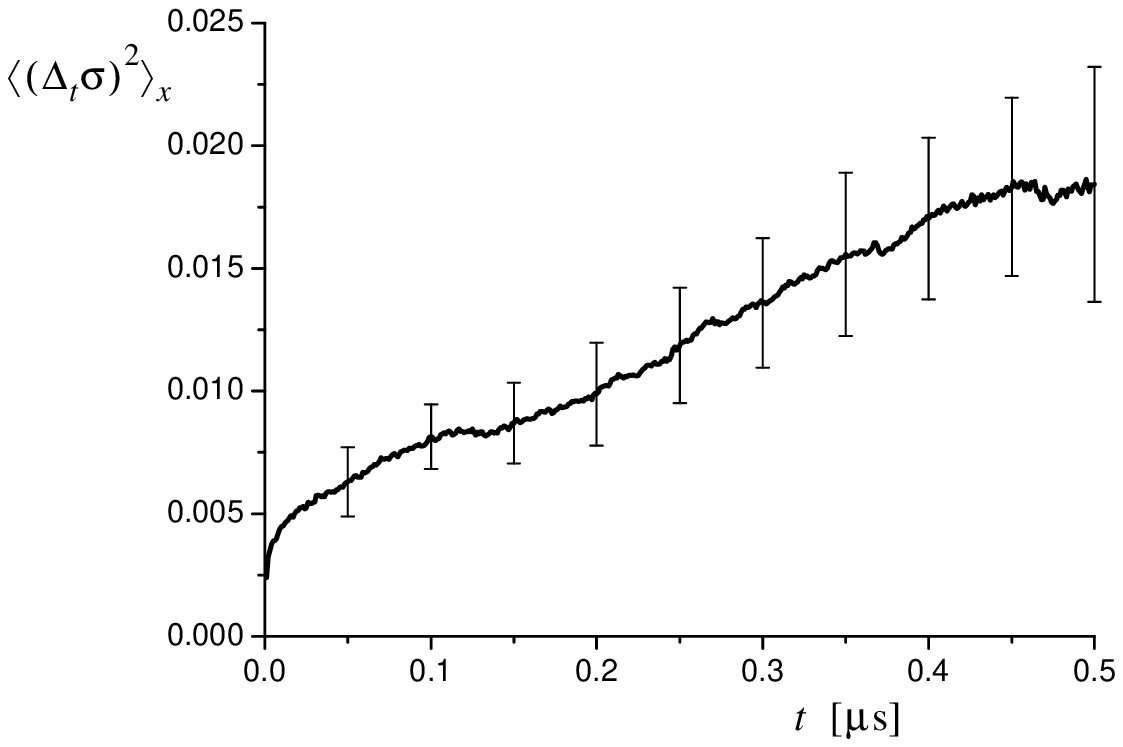}}
\vspace{1.5cm}

\noindent\emph{Figure 4. Diffusion along the primitive path
constructed for a polyethylene chain with $1000$ carbon atoms; the
slope represents the dynamic material information for the
reptation model. [Figure courtesy of M.~Kr\"oger, ETH Z\"urich.]}

\end{document}